\documentclass[aps,prd,nofootinbib,floatfix,preprintnumbers,amsmath,amssymb]{revtex4}
\usepackage{graphicx}
\usepackage{multirow}
\usepackage[english]{babel}
\usepackage{amsmath}
\usepackage{graphicx}
\usepackage{color}
\usepackage{amssymb}
\usepackage{hyperref}
\usepackage{dcolumn}% Align table columns on decimal point
\usepackage{bm}% bold math

\begin{document}

\title{Yukawa textures or dark doublets from Two Higgs Doublet Models with $Z_3$ symmetry }

\author{
Alfredo Aranda,$^{1,2}$\footnote{Electronic address: fefo@ucol.mx}
J. Hern\'andez--S\'anchez,$^{2,3}$\footnote{Electronic address:  jaime.hernandez@correo.buap.mx}
Roberto Noriega-Papaqui$^{2,4}$\footnote{Electronic address: rnoriega@uaeh.edu.mx}
Carlos A. Vaquera-Araujo,$^{1}$\footnote{Electronic address: carolusvaquera@gmail.com}
\vspace*{0.3cm}}

\affiliation{$^1$Facultad de Ciencias, CUICBAS,
Universidad de Colima, Colima, M\'exico \\
$^2$Dual C-P Institute of High Energy Physics, M\'exico \\
$^3$ Fac. de Cs. de la Electr\'onica, Benem\'erita Universidad Aut\'onoma de Puebla, Apdo. Postal 542, 72570 Puebla, Puebla, M\'exico \\
$^{4}$ \'Area Acad\'emica de Matem\'aticas y F\'{\i}sica, Universidad Aut\'onoma del Estado
de Hidalgo, Carr. Pachuca-Tulancingo Km. 4.5, C.P. 42184, Pachuca, Hgo.}

\date{\today}

\begin{abstract}
The effect of $Z_3$ symmetry on the general Two Higgs Doublet Model is explored. Of particular interest is the question of what can a $Z_3$ symmetry do beyond the usual case with $Z_2$. There are two independent scenarios that give some interesting results: first, by giving non-trivial charges to the Standard Model fermions, it is possible to use the $Z_3$ symmetry of the scalar potential to generate potentially useful Yukawa textures. This is not possible with $Z_2$. A series of possibilities is presented where their viability is addressed and a specific example in the quark sector is given for concreteness. The second venue of interest is in the area of inert doublets. It is shown that by considering the Standard Model plus two additional inert doublet scalars, i.e. a Dark Two Higgs Doublet Model, together with $Z_3$, a scenario can be obtained that differs from the $Z_2$ case. Some general comments are presented on the potentially interesting phenomenology of such model.
\end{abstract}
\maketitle
\section{Introduction}
Two Higgs Doublet Models (2HDMs) have been studied for a long time. They represent the next non-trivial step in complication beyond the Standard Model (SM) and yet, the simple addition of one scalar doublet, enriches the phenomenology substantially, leading to a wide spectrum of interesting phenomenological and theoretical possibilities~\cite{Gunion:1989we,Branco:2011iw}.

From the beginning it was realized that discrete symmetries play an important role in the scalar potential. In particular $Z_2$ symmetry has been widely used to restrict the general 2HDM potential leading to interesting and potentially relevant phenomenological consequences. In this letter we explore the general 2HDM with $Z_3$ symmetry and point out some interesting scenarios that could behold a phenomenological interest. In particular we are interested in finding features not present in the $Z_2$ symmetry case that might lead to interesting models. The first and strongest realization of this comes from the fact that the $Z_3$ symmetry may also affect the fermions fields in a non-trivial way, as opposed to the $Z_2$ case, leading to peculiar Yukawa textures~\cite{Ferreira:2010ir}. We find that in fact this happens in the quark sector for a class of models where one of the doublets is restricted to couple to the second and third quark family, while the other doublet couples to everything. Given appropriate $Z_3$ charges for the different quark fields and scalar doublets, acceptable Yukawa textures that lead to the correct Cabibbo-Kobayahi-Maskawa (CKM) matrix can be obtained.  

Another interesting and timely venue is that of the so-called inert models, i.e. scenarios where there can be scalar doublets with vacuum expectation values (vevs) equal to zero and a discrete symmetry, usually $Z_2$, that prevents the lightest one from decaying~\cite{Deshpande:1977rw,Barbieri:2006dq}. Since the vevs of these scalar fields are zero, they do not contribute to symmetry breaking. Furthermore, their charges (and that of the fermions in the SM) are chosen so that they do not couple to SM fermions - thus they are {\it inert}. In this case we address the issue of whether or not $Z_3$ symmetry can be used to stabilize a dark matter candidate and if a difference from the $Z_2$ case can be obtained. The answer to both questions is positive and we present the specific case where that happens. It consists on an extension to three scalar doublets: one active and a dark-2HDM with $Z_3$ symmetry.

In section~\ref{sec:analysis} we present the general setup for the scalar potential and the analysis in the Yukawa sector. Section~\ref{sec:yukawatextures} contains different sets of possible models with their generated Yukawa textures. The different inert type scenarios are presented in section~\ref{sec:inert} and we end with some final comments in section~\ref{sec:conclusion}.

\section{Analysis} 
\label{sec:analysis}
\subsection{General scalar sector}
There are two SU(2) doublets $H_i$, $i=1,2$, with the same hypercharge. We impose a $Z_3$ symmetry under which an arbitrary field field ${\cal F}$ transforms as ${\cal F} \rightarrow  {\cal F}^{\prime} = \omega^{n_f} {\cal F}$, where $\omega \equiv \exp(2\pi i/3)$ and $|n_f| \in \{0,1,2\}$. We say the field ${\cal F}$ has {\it charge} $n_f$ under $Z_3$. For simplicity we assume CP conservation in the scalar sector and thus all vevs and couplings are taken to be real.

The general SU(2)$_W\times$U(1)$_Y$ invariant scalar potential can be written as
\begin{eqnarray} \label{potential}
V(H) = \mu_{ij}^2 H_i^{\dagger}H_j + \lambda_{ijkl}\left( H_i^{\dagger}H_jH_k^{\dagger}H_l + h.c.\right) \ ,
\end{eqnarray}
where $i,j = 1,2$. Denoting the $Z_3$ charges of $H_1$ and $H_2$ as $n_{h_1}$ and $n_{h_2}$ respectively we can express the $Z_3$ charges of each term in the potential using the coefficients $\mu$ and $\lambda$. For example the coefficients of the (always) invariant quadratic terms involving $H_1$ or $H_2$ (but not their mixing) are said to have zero charge: $ [\mu_{ii}] = 0$. The other invariant terms are  $\left[ \lambda_{iiii}\right]  = \left[ \lambda_{ijji}\right]   = 0$. The rest are obtained from the following relations $(\ i\neq j )$:
\begin{eqnarray}\label{potential-lambda-terms} 
[\mu_{ij}] &=& -[\mu_{ji}]=  [\lambda_{iiij}] = \frac{1}{2} [\lambda_{ijij}] = n_{h_j}-n_{h_i} \pmod{3}  \ ,\\ 
\left[\lambda_{ijii}\right] &=& \left[\lambda_{iiij}\right]  = -\left[\lambda_{jiii}\right]  = -\left[\lambda_{iiji}\right] \ .
\end{eqnarray}

Note that the particular case where $n_{h_j}-n_{h_i} \neq 0 \pmod{3}$ leads to a $Z_2$ invariant potential: it does not have $H_i^{\dagger}H_j$ quadratic terms and only six quartic terms survive. Also, If both scalars develop nonzero vevs, the gauge symmetry is completely broken and soft breaking terms must be included in order to obtain a $U(1)_{em}$ invariant vacuum~\cite{Gunion:1989we}. Thus we are interested in the situation where $n_{h_j}-n_{h_i} \neq 0 \pmod{3}$ and the possible consequences that this may have in the Yukawa sector. The case $n_{h_1}=n_{h_2}$ is possible but uninteresting.

\subsection{Yukawa sector}

The Yukawa sector of the general scenario is given by

\begin{eqnarray}\label{yukawa}
{\cal L}_Y = {\cal Y}^{ui}_{ab} \overline{Q}_a \widetilde{H}_i u_b + {\cal Y}^{di}_{ab} \overline{Q}_a H_i d_b + h.c. \ ,
\end{eqnarray}
where $i=1,2$, $a,b=1,2,3$, and $\widetilde{H}_i \equiv i\sigma_2H_i^*$.
Denoting the $Z_3$ charges of the different fields by $n_{h_i}$, $n_{q_a}$, $n_{u_a}$, $n_{d_a}$, the "charge" associated to the Yukawas become
\begin{eqnarray}
\label{yukawa-up-charges}
\left[  {\cal Y}^{ui}_{ab}   \right] & = & n_{u_b} - n_{q_a} - n_{h_i}  \pmod{3} \ , \\
\label{yukawa-down-charges}
\left[  {\cal Y}^{di}_{ab}   \right] & = & n_{d_b} - n_{q_a} + n_{h_i}   \pmod{3} \ . 
\end{eqnarray}

These expressions can be used to determine the type of Yukawa textures that can be obtained from different charge assignments. Our purpose in this work is to  present a few possibilities that can lead to the construction of  models and so, instead of making a full classification of all possible models (as has been done in~\cite{Ferreira:2010ir}), we explore some specific arrangements that we feel can be of phenomenological interest. We restrict the discussion to the quark sector for the moment (in order not to include the extra fields required to explore the lepton sector~\cite{Aranda:2011rt}) but stress that the possibilities presented here can be successfully extended, as will be shown in a future publication.

\section{Particular cases: Yukawa textures}
\label{sec:yukawatextures}
\subsection{One of up, one for down}
An interesting possibility is whether or not the $Z_3$ symmetry can be used for "separating" the up and down-type sectors. From Eqs.\eqref{yukawa-up-charges} and \eqref{yukawa-down-charges} it can be seen that this happens for the following case: given $n_{q_a}$ for $Q_a$, then $H_1$ couples to the up-type sector and $H_2$ to the down-type one if $n_{h_1} \neq n_{h_2}$, $n_{u_b}=n_{q_a}+n_{h_1}$, and $n_{d_b}=n_{q_a}-n_{h_2}$ ($ \pmod{3}$ is implied in all these relations).

Analyzing the different charge combinations leads to Yukawa matrices that are either diagonal or block-diagonal, i.e., the possible textures are
\begin{eqnarray}
\label{yukawas-diagonal}
{\cal Y} \sim 
\left( \begin{array}{ccc}
\star & 0 & 0 \\
0& \star & 0 \\
0&0&\star
\end{array}\right), \ \
{\cal Y} \sim 
\left( \begin{array}{ccc}
\star & \star & 0 \\
\star & \star & 0 \\
0&0&\star
\end{array}\right), \ \
{\cal Y} \sim 
\left( \begin{array}{ccc}
\star & 0 & 0 \\
0& \star & \star \\
0&\star&\star
\end{array}\right).
\end{eqnarray}
These textures are not useful as they do not lead to the observed CKM matrix~\cite{Ramond:1993kv,Roberts:2001zy}.

\subsection{Avoiding one and one for all} \label{oneforthirdoneforall}
Another possibility consists on having a doublet, say $H_1$, coupling to all three families while the other couples only to the $2-3$ sector (thus avoids one!). This happens when the following conditions are met:
\begin{eqnarray}\label{conditions-23} \nonumber
\left[n_{q_2} \right] &=& \left[n_{q_3} \right] , \ \left[n_{u_2} \right]  = \left[n_{u_3} \right], \ 
\left[n_{q_1} \right]  \neq \left[n_{u_3} \right] -\left[n_{h_2} \right], \\ \left[n_{u_1} \right]  &\neq& \left[n_{q_3} \right] +\left[n_{h_2} \right], \left[n_{u_1} \right]  \neq \left[n_{q_1} \right] +\left[n_{h_2} \right] \  \pmod{3}.
\end{eqnarray}

Given a value for $n_{h_2}$ there are six possible combinations that satisfy the conditions above. The 18 combinations are given in table~\ref{table:oneforallonefor3}.

\begin{center}
\begin{table}[ht]
\begin{tabular}{|ccccc||ccccc||ccccc|}
\hline
$n_{h_2}$ & $n_{u_3}$ & $n_{q_3}$ & $n_{u_1}$ & $n_{q_1}$  &
$n_{h_2}$ & $n_{u_3}$ & $n_{q_3}$ & $n_{u_1}$ & $n_{q_1}$  &
$n_{h_2}$ & $n_{u_3}$ & $n_{q_3}$ & $n_{u_1}$ & $n_{q_1}$ \\
\hline
\hline
0 & 0 & 0 & 1 & 2 & 1 & 0 & 2 & 1 & 1 & 2 & 0 & 1 & 1 & 0\\
0 & 0 & 0 & 2 & 1 & 1 & 0 & 2 & 2 & 0 & 2 & 0 & 1 & 2 & 2\\
{\bf 0} & {\bf 1} & {\bf 1} & {\bf 0} & {\bf 2} & 1 & 2 & 1 & 0 & 0 & 2 & 1 & 2 & 0 & 0\\
0 & 1 & 1 & 2 & 0 & 1 & 2 & 1 & 1 & 2 & 2 & 1 & 2 & 2 & 1\\
0 & 2 & 2 & 0 & 1 & 1 & 1 & 0 & 0 & 1 & 2 & 2 & 0 & 0 & 1\\
0 & 2 & 2 & 1 & 0 & 1 & 1 & 0 & 2 & 2 & 2 & 2 & 0 & 1 & 2\\
\hline
\end{tabular}
\caption{Given a charge for $n_{h_2}$, this table lists the 18 possible charge combinations consistent with Eqs.\eqref{conditions-23} for all remaining fields. The case in bold case is used later in the text, this is why it has been singled out.}
\label{table:oneforallonefor3}
\end{table}
\end{center}
 
As an example consider the (bold case) case in the third row of the first column of Table~\ref{table:oneforallonefor3}. It corresponds to the following charge assignments: $n_{u_1}=0$, $n_{u_2}=n_{u_3}=1$, $n_{q_1}=2$, $n_{q_2}=n_{q_3}=1$, and $n_{h_2}=0$. Then, writing the up-type quark mass matrix as ${\cal Y}^u \equiv {\cal Y}^{u1} + {\cal Y}^{u2}$, we get
\begin{eqnarray}
{\cal Y}^{u2} = 
\left( \begin{array}{ccc}
0 & 0 & 0 \\
0& \star & \star \\
0&\star&\star
\end{array}\right), \ \ 
{\cal Y}^{u1}[n_{h_1}=1] = 
\left( \begin{array}{ccc}
0 & 0 & 0 \\
0 & 0 & 0 \\
0 & 0 & 0
\end{array}\right), \ \ 
{\cal Y}^{u1}[n_{h_1}=2]  = 
\left( \begin{array}{ccc}
0 & * & * \\
* & 0 & 0 \\
* & 0 & 0
\end{array}\right), \ \ 
\end{eqnarray}
and so the two possibilities
\begin{eqnarray}\label{finalUyukawas-case1}
{\cal Y}^u[n_{h_1}=1] =
\left( \begin{array}{ccc}
0 & 0 & 0 \\
0 & \star & \star \\
0 & \star & \star
\end{array}\right), \ \ 
{\cal Y}^u[n_{h_1}=2] =
\left( \begin{array}{ccc}
0 & * & * \\
* & \star & \star \\
* & \star & \star
\end{array}\right).
\end{eqnarray}

In these expressions the symbols $\star$ and $*$ represent products of unknown dimensionless coefficients with the vevs of $H_2$ and $H_1$ respectively. The second possibility in Eq.~\eqref{finalUyukawas-case1}, corresponding to $n_{h_1}=2$, is clearly viable (of course one now has to determine the down-type sector as well as leptons) while the first one does not work.

Lets check the down-type sector. Concentrating on the second case, i.e. the one corresponding to $n_{h_1}=2$, and using Eq.~\eqref{yukawa-down-charges}, we obtain (expressing the charges of each entry):
\begin{eqnarray}\label{Dyukawa-case1} \nonumber
\left[{\cal Y}^{d1}[n_{h_1}=2]\right] = \left[
\left( \begin{array}{ccc}
n_{d_1} & n_{d_2} &n_{d_3} \\
n_{d_1}+1 & n_{d_2} +1&  n_{d_3} +1 \\
n_{d_1}+1 & n_{d_2}+1 &  n_{d_3} +1
\end{array}\right) \right], \ \
\left[{\cal Y}^{d2}[n_{h_1}=2]\right] = \left[
\left( \begin{array}{ccc}
n_{d_1} -2 & n_{d_2} -2 &n_{d_3} -2 \\
n_{d_1}-1 & n_{d_2} -1&  n_{d_3} -1 \\
n_{d_1}-1 & n_{d_2}-1 &  n_{d_3} -1
\end{array}\right)\right]. \\
\end{eqnarray}

Now lets choose as an example the following charge assignments: $n_{d_1}=1$ and $n_{d_2}=n_{d_3}=2$. Thus, we obtain
\begin{eqnarray}
{\cal Y}^{d2} = 
\left( \begin{array}{ccc}
0 & \star & \star \\
\star & 0 & 0 \\
\star & 0 & 0
\end{array}\right), \ \ 
{\cal Y}^{d1}[n_{h_1}=2] = 
\left( \begin{array}{ccc}
0 & 0 & 0 \\
0 & * & * \\
0 & * & *
\end{array}\right), 
\end{eqnarray}
leading to
\begin{eqnarray}\label{finalDyukawa-case1}
{\cal Y}^d[n_{h_1}=2] =
\left( \begin{array}{ccc}
0 & \star & \star \\
\star & * & * \\
\star & * & *
\end{array}\right).
\end{eqnarray}

Note that the role of $H_1$ and $H_2$ is inversely correlated between ${\cal Y}^u$ and ${\cal Y}^d$ in Eqs.~\eqref{finalUyukawas-case1} and~\eqref{finalDyukawa-case1}.

This example is one of several that can lead to interesting Yukawa textures in agreement with the CKM matrix. A complete phenomenological analysis of these models, including the lepton sector, is beyond the scope of this letter and will be presented in a future publication.

\subsection{One up and one for all}
One can also look for a case where one of the doublets (say $H_2$) couples to the up-type sector only while the other couples to everything. From Eq.~\eqref{yukawa-down-charges} we obtain the restriction $n_{d_b} \neq n_{h_2} + n_{q_a} \pmod{3}$. There are three possibilities: i) all $n_{q_a}$ are different and thus we cannot satisfy the condition, ii) two equal, i.e. WLOG $n_{q_1} = n_{q_2} \neq n_{q_3}$, then all three $n_{d_a}$ must be equal leading to unacceptable Yukawa textures. Finally iii) where the three $n_{q_a}$ are equal. This last case leads to Yukawa matrices with either no zeros or complete columns equal to zero. These textures are not useful.

\section{Inert models}
\label{sec:inert}

The inert model corresponds to the situation where all fields have zero charge under $Z_3$ except for one of the scalar doublets, say $H_2$, and only $H_1$ gets a nonzero vev. In this case the $Z_3$ symmetry is not broken and there is a stable neutral particle in the spectrum. Note that this case would correspond to the situation described right after Eq.~\eqref{potential-lambda-terms} where $n_{h_2}-n_{h_1}\neq 0$, except that the correct breaking of the gauge symmetry is accomplished without any soft breaking terms.

The potential in this case is given by
\begin{eqnarray} \label{inertpotential} \nonumber
V(H_1 ,H_2) & = & \mu^2_1 H_1^{\dagger}H_1+\mu^2_2 H_2^{\dagger}H_2 + \lambda_1 \left( H_1^{\dagger}H_1 \right)^2  +\lambda_2 \left( H_2^{\dagger}H_2 \right)^2 + \\ 
& + & \lambda_3 \left( H_1^{\dagger}H_1 \right)\left( H_2^{\dagger}H_2 \right) + \lambda_4 \left( H_1^{\dagger}H_2 \right)\left( H_2^{\dagger}H_1 \right) \ ,
\end{eqnarray}
and it corresponds to the general (inert) case with arbitrary $Z_n$ (except for $n=2$ where the term $\mathfrak{Re} \left( (H_1^{\dagger}H_2)^{2}\right)$ is also present). The $Z_3$ case does not contribute anything new.

A potentially interesting extension is to consider a third Higgs doublet (for a similar case with $Z_2$ see~\cite{Keus:2014jha} and for recent work where multi-inert doublets are explored see~\cite{Keus:2014isa}). There are two possibilities: either two of them are neutral under $Z_3$ and the third one does not acquire a vev (hence this third one would contain the dark matter candidate) or just one of the doublets is neutral and gets a vev with an additional "Dark Two Higgs Doublet model" D2HDM (note that this is different from the model presented in~\cite{Lee:2013fda}, where the authors used the same name for a model with a normal 2HDM plus an inert singlet scalar.). Of course, in either case, the $Z_3$ symmetry does not play any role in the Yukawa sector and would be useful only in order to stabilize the dark matter candidate. The idea would be to determine if using $Z_3$ would give something different from the $Z_2$ case.

Lets consider the first case and call it the Inert 3HDM. We introduce a third scalar doublet $H_3$ charged under $Z_3$ with charge $n_{h_3}$ and let $H_1$ and $H_2$ be neutral. The potential is given by
\begin{eqnarray} \label{I3HDMpotential} \nonumber
V(H_i) & = &  \mu^2_i H_i^{\dagger}H_i + \lambda_i \left( H_i^{\dagger}H_i \right)^2  + \lambda_{\substack{ij \\
        i < j}} \left( H_i^{\dagger}H_i \right)\left( H_j^{\dagger}H_j \right)  +  \lambda^{\prime}_{\substack{ij \\
        i < j}} \left( H_i^{\dagger}H_j \right)\left( H_j^{\dagger}H_i\right)  + \\ \nonumber
& + & \left\{ \lambda_4 \left( H_3^{\dagger}H_3 \right)\left( H_1^{\dagger}H_2 \right) + \lambda_{123} \left( H_1^{\dagger}H_3 \right)\left( H_3^{\dagger}H_2 \right) 
+ \lambda^{\prime\prime}_{\substack{ab \\
        a < b}}  \left( H_a^{\dagger}H_b \right)\left( H_b^{\dagger}H_b \right)  + h.c. \right\}, \\
\end{eqnarray}
where $i,j=1,2,3$ and $a,b=1,2$. Again, this potential is quite general and not specific to $Z_3$. It  is however interesting and can, in principle, contain regions of parameter space where a viable dark matter candidate might exist. In order to determine that one must check all appropriate conditions. At this moment we leave it open as a possibility and jump to the other, namely, the D2HDM where there are two scalar SU(2) doublets charged under $Z_3$ and one neutral. 

Let $H_1$ denote the $Z_3$ neutral scalar. Then its vev breaks electroweak symmetry but leaves $Z_3$ unbroken. The potential in this case is more constrained and the number of terms depends on whether $H_2$ and $H_3$ have the same non-zero $Z_3$ charge or not:
\begin{eqnarray} \label{D2HDMpotential} \nonumber
V(H_i) & = &  \mu^2_i H_i^{\dagger}H_i + \lambda_i \left( H_i^{\dagger}H_i \right)^2  + \lambda_{\substack{ij \\
        i < j}} \left( H_i^{\dagger}H_i \right)\left( H_j^{\dagger}H_j \right)  +  \lambda^{\prime}_{\substack{ij \\
        i < j}} \left( H_i^{\dagger}H_j \right)\left( H_j^{\dagger}H_i\right)  + \\ \nonumber
& + & \left[ \lambda^{\prime \prime}_{ab} \left( H_1^{\dagger}H_a \right)\left( H_b^{\dagger}H_a \right)  + h.c. \right]+ 
 \left\{ \mu^2_{23} H_2^{\dagger}H_3  + \lambda_{123} \left( H_1^{\dagger}H_1 \right)\left( H_2^{\dagger}H_3 \right) 
+ h.c. \right\}, \\
\end{eqnarray}
with $a = 2, 3$ and where the terms inside the curly brackets survive in the case where $H_2$ and $H_3$ have the same $Z_3$ charge, and so, if they do not, the potential without those terms is different from the one that we would obtained using $Z_2$ instead. . 

The interesting (different from $Z_2$) case then corresponds to $n_{h_3}\neq n_{h_2}$ leading to the $Z_3$ invariant potential
\begin{eqnarray} \label{D2HDMZ·potential} \nonumber
V(H_i) & = &  \mu^2_i H_i^{\dagger}H_i + \lambda_i \left( H_i^{\dagger}H_i \right)^2  + \lambda_{\substack{ij \\
        i < j}} \left( H_i^{\dagger}H_i \right)\left( H_j^{\dagger}H_j \right)  \\ 
& + &   \lambda^{\prime}_{\substack{ij \\
        i < j}} \left( H_i^{\dagger}H_j \right)\left( H_j^{\dagger}H_i\right)  +
 \left[ \lambda^{\prime \prime}_{ab} \left( H_1^{\dagger}H_a \right)\left( H_b^{\dagger}H_a \right) + h.c. \right] \ .
\end{eqnarray}

The "dark" phenomenology of this model can be substantially different from the case with one inert doublet. This D2HDM has two charged dark scalars, two dark pseudoscalars and two CP-even dark scalars. Depending on the mass spectrum for the dark sector there will be interesting cascade and radiative processes not present in the case where the is only on inert doublet~\cite{Keus:2014isa}. A complete study of the properties and signals for this model along the lines of~\cite{Keus:2014isa} will be presented in a forthcoming publication.

\section{Conclusion}
\label{sec:conclusion}
We have explored some consequences of imposing $Z_3$ symmetry to the scalar potential of the Two Higgs Doublet Model. The exploration went into two main venues: on one hand, by charging the Standard Model fermions under the $Z_3$, it is possible to generate interesting Yukawa textures, something that the usual case with $Z_2$ symmetry cannot do. We have presented some cases that can lead to phenomenological viable models, namely cases that can reproduce the CKM matrix and, if extended as in~\cite{Aranda:2011rt}, the PMNS matrix in the lepton sector. The successful case corresponds to a situation where one of the doublets is free to couple to all generations whereas the other is restricted to couple to the $2-3$ sector only. Depending on the specific charges, several models can emerge with suitable Yukawa textures and a full phenomenological analysis of those models will be presented elsewhere. We discussed other two scenarios that however did not succeed in providing useful Yukawa textures: i) requiring one doublet to couple to the up-type sector and the other to the down-type sector and ii) one doublet free to couple to everything while the other restricted to the up-type sector. 

The second venue of exploration consisted in the implementation of the so-called inert models within the framework of $Z_3$. In this case the doublet(s) that can get a vev must be neutral under $Z_3$ (or any other symmetry that might be used in the scalar sector) and so after electroweak symmetry breaking the $Z_3$ si preserved. If there is a doublet (or several) that is charged nontrivially under the $Z_3$ (and thus is not allowed to have a vev), then there is a stable particle in the spectrum that can in principle become a dark matter candidate. We showed that a particularly promising scenario is the one in which there is the usual Standard Model Higgs doublet plus two inert doublets with $Z_3$ symmetry, a Dark Two Higgs Doublet Model. This scenario can be different from the one with $Z_2$ symmetry and can lead to interesting phenomenology. A full analysis of this model is under investigation.

\acknowledgments
This work was supported in part by CONACYT (M\'exico) and by PROMEP (M\'exico) under the grant: Red Tem\'atica: "F\'isica del Higgs y del sabor."

\bibliographystyle{ieeetr}

\end{document}